\documentclass[12pt]{article}
\usepackage{axodraw}
\newcommand{\beq}{\begin{equation}}
\newcommand{\eeq}{\end{equation}}
\newcommand{\beqa}{\begin{eqnarray}}
\newcommand{\eeqa}{\end{eqnarray}}
\newcommand{\beqar}{\begin{eqnarray*}}
\newcommand{\eeqar}{\end{eqnarray*}}

\newcommand{\eps}{\epsilon}

\newcommand{\Ga}{\Gamma}

\newcommand{\inn}{\!\cdot\!}

\newcommand{\lam}{\lambda}

\newcommand{\z}{\zeta}

\newcommand{\eg}{{\it e.g.,}\ }
\newcommand{\ie}{{\it i.e.,}\ }
\newcommand{\labell}[1]{\label{#1}} %{\label{#1}} %
\newcommand{\reef}[1]{(\ref{#1})}
\newcommand\prt{\partial}

\newcommand\cL{{\cal L}}

\newcommand\cD{{\cal D}}
\newcommand\cT{{\cal T}}

\newcommand\bD{\bar{D}}

\newcommand\Tr{{\rm Tr}}
\newcommand\STr{{\rm STr}}
\begin{document}

\thispagestyle{empty} \rightline{\small  \hfill }
\vspace*{1cm}

\begin{center}
{\Large \bf On  higher derivative  corrections of  tachyon  action }
\vspace*{1cm}

{Mohammad R. Garousi$^{a,b}$ and Hanif Golchin$^a$}\\
\vspace*{1cm}
{ ${}^a$Department of Physics, Ferdowsi university, P.O. Box 1436, Mashhad, Iran}\\
\vspace*{0.1cm}

{ ${}^b$School of Physis, IPM (Institute for Studies in Theoretical Physics and Mathematics),} \\
{P.O. Box 19395-5531, Tehran, Iran}\\
\vspace*{0.4cm}

\vspace{2cm}
ABSTRACT
\end{center}
 We  have examined the momentum expansion of the disk level S-matrix element of two tachyons and two gauge fields to find, up to on-shell ambiguity, the couplings of these fields in the world volume theory of N coincident  non-BPS D-branes to all order of $\alpha'$. Using the  proposal  that the  action of D-brane-anti-D-brane is given by the projection of the  action of two non-BPS D-branes with $(-1)^{F_L}$, we find the corresponding couplings  in the world volume theory of the brane-anti-brane system. Using these infinite tower of couplings, we then calculate the massless pole of the scattering amplitude of one RR field, two tachyons and one gauge field in the brane-anti-brane theory.  We find that the massless pole of the field theory amplitude is exactly equal to  the massless pole of the disk level S-matrix element of one RR, two tachyons and one gauge field  to all order of $\alpha'$. 
%The above couplings indicate that the two tachyons and two gauge fields couplings of the non-abelian tachyon DBI %action are the effective couplings when gauge field varies slowly.\\
We  have also  found the couplings of four tachyons   to all order of $\alpha'$ by examining the  S-matrix element of four tachyons. 
%These couplings indicate that for $N=1$ case, the four tachyons couplings of the tachyon DBI action are the effective %coupling when tachyon varies slowly. 

\vfill
\setcounter{page}{0}
\setcounter{footnote}{0}
\newpage

\section{Introduction} \label{intro}
Brane-anti-brane system has been used to model inflation in string theory\cite{Dvali:1998pa,Alexander:2001ks,Dvali:2001fw,Kachru:2003sx}. When branes are very far away from each other, the transverse scalar field which describes the motion of one brane in the background of the other brane plays the role of inflaton. The dynamics of the moving brane  in this period is very well described at low energy by the DBI action. On the other hand, when branes come within a critical distance from each other,  the string stretching between the two branes become tachyonic and inflation ends.   The dynamics of the brane-anti-brane in this period is very important for studying the reheating \cite{Traschen:1990sw,Kofman:1997yn,Brandenberger:2007ca}.  
Brane-anti-brane system  has been also used to study spontaneous chiral symmetry breaking  in  holographic model of QCD \cite{Casero:2007ae,Bergman:2007pm,Dhar:2007bz}. In these studies, flavor branes introduced by placing a set of parallel branes and anti-branes on a background dual to a confining color theory \cite{Sakai:2004cn}. 

It is important then to study the world-volume theory of D$_p$-brane-anti-D$_p$-brane. The world-volume theory of this system has tachyon, massless  and infinite tower of massive fields which can be described by   Berkovits'
 superstring field theory \cite{Berkovits:1995ab,Berkovits:1998bt}. The world-volume theory may be rewritten in terms of the tachyon and massless fields and infinite number of derivative terms reflecting the effect of massive fields. We  call this field theory `` the higher derivative theory''. When the world volume fields vary slowly the higher derivative theory should be reduced to the effective theory. It is known that the  vortex solution of the field theory of the D$_p$-brane-anti-D$_p$-brane pair should describe the  stable D$_{p-2}$-brane \cite{Sen:1998tt}.  An effective action for brane-anti-brane  which has this property has been proposed in  \cite{Sen:2003tm} 
\beqa
S&=&-T_p\int d^{p+1}\sigma V(T)\left(\sqrt{-\det\textbf{A}^{(1)}}+\sqrt{-\det
\textbf{A}^{(2)}}\right)\,,\labell{action21}\eeqa where \beqa
\textbf{A}_{\mu\nu}^{(n)}&=&\eta_{\mu\nu}
+2\pi\alpha'F^{(n)}_{\mu\nu}+
\pi\alpha'\left(D_{\mu}T(D_{\nu}T)^*+D_{\nu}T(D_{\mu}T)^*\right)\,.\eeqa
where  $T_{p}$ is the
D$_p$-brane tension and $D_{a}T=\prt_{a}T-i(A^{(1)}_a-A^{(2)}_a)T$. The above action is a generalization of the tachyon DBI action \cite{Sen:1999md,Garousi:2000tr,Bergshoeff:2000dq,Kluson:2000iy}. 
The above action has a vortex solution whose world-volume action is given by the DBI action of  stable $D_{p-2}$-brane \cite{Sen:2003tm}. It is  difficult to find the higher derivative corrections to this action.

Another proposal for the effective action of the brane-anti-brane pair which is based on the S-matrix elements calculation is given by \cite{Garousi:2004rd,Garousi:2007fn}
\beqa
S_{DBI}&=&-T_p\int
d^{p+1}\sigma \STr\left(V({\cal T})
\sqrt{-\det(\eta_{ab}
+2\pi\alpha'F_{ab}+2\pi\alpha'D_a{\cal T}D_b{\cal T})} \right)\,\,,\labell{nonab} \eeqa  The trace in the above action 
should be completely symmetric between all  matrices
of the form $F_{ab},D_a{\cal T}$, and individual
${\cal T}$ of the tachyon potential.   These matrices  are
\beqa
F_{ab}=\pmatrix{F^{(1)}_{ab}&0\cr 
0&F^{(2)}_{ab}},\,\,
D_{a}{\cal T}=\pmatrix{0&D_aT\cr 
(D_aT)^*&0},\,\, {\cal T}=\pmatrix{0&T\cr 
T^*&0}\,\labell{M12} \eeqa 
To implement the symmetric trace prescription, one must first expand the action then make each terms symmetric and finally take the trace. This in particular gives a coupling between $F^{(1)}$ and $F^{(2)}$. There is no such coupling in \reef{action21}. The above action has been found  from the effective field theory of $N=2$ non-BPS branes by projecting it with $(-1)^{F_L}$ where $F_L$ is the spacetime left-handed fermion number. On the other hand, the effective field theory of two non-BPS D-branes has been  assumed to be the natural non-abelian extension of the tachyon DBI action, \ie the action \reef{nonab} without restricting the matrices to \reef{M12}. In this paper, we would like to study the higher derivative corrections to this action.

A method for finding the higher derivative theory is to study the S-matrix elements of this theory and compare them with the S-matrix elements of string theory. If this higher derivative  theory is going to be identical with the string theory, the S-matrix elements of the higher derivative theory must be identical to the momentum expansion of the  S-matrix elements of string theory. Hence, by calculating the S-matrix elements of string theory and finding their momentum expansions, one can find the appropriate higher derivative couplings in the field theory.  For instance, the string theory S-matrix element of one RR field and two tachyons can be reproduced by the higher derivative theory of the brane-anti-brane system  if it includes the following couplings \cite{Garousi:2007si}: 
\beqa
2i\alpha'\mu_p\sum_{n=0}^{\infty}a_n\left(\frac{\alpha'}{2}\right)^n C_{p-1}\wedge (D ^aD_a)^n(DT\wedge DT^*)\labell{hderv}
\eeqa
where $a_n$'s are some known numbers. Or the S-matrix element of one RR and two gauge fields can be reproduced by the higher derivative field theory if it includes
\beqa
\frac{\mu_p}{2!}(2\pi\alpha')^2C_{p-3}\wedge \left(\sum_{n=-1}^{\infty}b_n(\alpha')^{n+1}\partial^{a_1}\cdots\partial^{a_{n+1}} F\wedge \partial_{a_1}\cdots\partial_{a_{n+1}} F\right)\labell{highaa}\eeqa
where $b_n$'s  are some known numbers. In above couplings $F$ is one of the gauge fields of the brane-anti-brane system. In the above examples, it is trivial to find the momentum expansions of the S-matrix elements in the string theory side, however, in  higher point functions, it is nontrivial to find their momentum  expansions. The momentum expansion of the S-matrix element of one RR, two tachyons and one gauge field has been found in \cite{Garousi:2007fk,Garousi:2007si}. They can be reproduced by the higher derivative theory if it includes the following couplings for $C_{p-3}$: 
\beqa
&&2i\alpha'(\pi\alpha')\mu_p\sum_{p,n,m=0}^{\infty}c_{p,n,m}\left(\frac{\alpha'}{2}\right)^{p}\left(\alpha'\right)^{2n+m} C_{p-3}\wedge \prt^{a_1}\cdots\prt^{a_{2n}}\prt^{b_1}\cdots\prt^{b_{m}}F\nonumber\\
&&\wedge(D^aD_a)^p  D_{b_1}\cdots D_{b_{m}}(D_{a_1}\cdots D_{a_n}DT\wedge D_{a_{n+1}}\cdots D_{a_{2n}}DT^*)\labell{hderv12}
\eeqa
where again $c_{p,n,m}$'s  are some known numbers. And  the following couplings for $C_{p-1}$:
\beqa
-2\alpha'\mu_p\sum_{n=0}^{\infty}a_n\left(\frac{\alpha'}{2}\right)^n C_{p-1}\wedge (D^aD_a)^n(F|T|^2)\labell{hderv2}
\eeqa
where $a_n$'s  are exactly the numbers that appear in \reef{hderv}, and
\beqa
&&2(\alpha')^2\mu_p
\sum_{p,n,m=0}^{\infty}e_{p,n,m}(\alpha')^{2m+n}\left(\frac{\alpha'}{2}\right)^pC_{p-1}\wedge\labell{highpn'}\\
&& (D_aD^a)^p \left[-\prt_b\prt_c \prt^{a_1}\cdots \prt^{a_n}\prt_{b_1}\cdots\prt_{b_{2m}}FD_{a_1}\cdots D_{a_n}(D^bD^{b_1}\cdots D^{b_m} TD^c D^{b_{m+1}}\cdots D^{b_{2m}}T^*)\right.\nonumber\\&&\left.+2D_{a_1}\cdots D_{a_n}(D_bD^{b_1}\cdots D^{b_m} DT\wedge D_c D^{b_{m+1}}\cdots D^{b_{2m}}DT^*)\prt^{a_1}\cdots \prt^{a_n}\prt_{b_1}\cdots\prt_{b_{2m}}F^{bc}\right.\nonumber\\
&&\left.+\prt_b \prt^{a_1}\cdots \prt^{a_n}\prt_{b_1}\cdots\prt_{b_{2m}}F_{c}\wedge D_{a_1}\cdots D_{a_n}(D^b D^{b_1}\cdots D^{b_m}DTD^c D^{b_{m+1}}\cdots D^{b_{2m}}T^*)\right.\nonumber\\&&\left.+\prt_b \prt^{a_1}\cdots \prt^{a_n}\prt_{b_1}\cdots\prt_{b_{2m}}F_{c}\wedge D_{a_1}\cdots D_{a_n}(D^bD^{b_1}\cdots D^{b_m} DT^*D^c D^{b_{m+1}}\cdots D^{b_{2m}}T) \right]\nonumber\eeqa
 where $e_{p,n,m}$ are some other known numbers. It has been argued in \cite{Garousi:2007si} that the tachyon couplings in \reef{hderv}, \reef{hderv12} and \reef{hderv2} have no on-shell ambiguity. Having  different couplings in the higher derivative theory without on-shell ambiguity, one can then find the effective theory by restricting the fields to be slowly varying fields. It is shown in \cite{Garousi:2007si} that the above couplings reduce to the Wess-Zumion effective couplings of brane-anti-brane system \cite{Kennedy:1999nn,Kraus:2000nj,Takayanagi:2000rz} for slowly varying fields.
In this paper, we would like to  extend the above discussion to find the higher derivative tachyon  couplings corresponding to the non-abelian tachyon  DBI action.  
%It seems that the higher derivative couplings that we will find can reproduce the infinite massless and tachyonic %poles of the mometum expansion of string theory S-matrix elements, however, they have on-shell ambiguity, \ie $\cT\sim %2\alpha'\prt_a\prt^a\cT$. Hence, unlike the WZ part, one can not read the effective couplings  from the higher %derivative couplings of tachyon presented in this paper. 

An outline of the rest of paper is as follows. In the next section,  we find the momentum expansion  of the string theory S-matrix element of  two tachyons and two gauge fields in the world volume theory of $N$ non-BPS D-branes. We then write a tower of infinite number of two-tachyon-two-gauge field couplings which reproduce the above momentum expansion. We repeat the same steps to find the couplings of four massless transverse scalar fields and the couplings of four tachyons. In section 3, using the proposal that the action of brane-anti-brane can be  found  from the action of $N=2$ non-BPS branes by projecting it with $(-1)^{F_L}$, we find the corresponding couplings in the brane-anti-brane theory. Using the two-tachyon-two-gauge field couplings of brane-anti-brane, we calculate the massless pole of the scattering amplitude of one RR field, two tachyons and one gauge field. We then compare the result  with the corresponding massless pole in the  string theory S-matrix element \cite{Garousi:2007si}. We find exact agreement.   

\section{Higher derivative terms of non-BPS branes}

The world volume of $N$ coincident non-BPS D-branes has $N^2$ tachyons and $N^2$ gauge fields. The higher derivative theory of non-BPS branes  which includes the kinetic terms and the couplings of these fields may be found by studying the S-matrix elements  in the field theory and in the string theory. The S-matrix elements on the world volume of unstable branes may have no clear physical interpretation, however, one expects that the string theory S-matrix elements  should be reproduced by the higher derivative  theory if the two theories are going to be identical.  In this section we would like to find the  two-tachyons-two-gauge fields couplings and  four-tachyons  couplings  which produce the string theory S-matrix elements to all order of $\alpha'$. We begin with the S-matrix element of two tachyons and two gauge fields.

\subsection{Two tachyons and two gauge fields couplings}  

The S-matrix element of two gauge fields and two tachyons  in string theory side  is given by \cite{mgjs,jp}  \beqa A&=&4i(2\pi\alpha')T_p\left[\frac{1}{2}\z_1\inn\z_2 \left(-\alpha
\frac{\Ga(-2s)\Ga(1/2-2t)}{\Ga(-1/2-2s-2t)}-\beta
\frac{\Ga(-2s)\Ga(1/2-2u)}{\Ga(-1/2-2s-2u)}\right.\right.\nonumber\\
&&\left.\left.+\gamma
\frac{\Ga(1/2-2u)\Ga(1/2-2t)}{\Ga(1+2s)}\right)\right.\nonumber\\
&&\left.+2\alpha'\z_1\inn k_3\z_2\inn k_4\left(\alpha
\frac{\Ga(-2s)\Ga(1/2-2t)}{\Ga(1/2-2s-2t)}-\beta
\frac{\Ga(-2s)\Ga(-1/2-2u)}{\Ga(-1/2-2s-2u)}\right.\right.\nonumber\\
&&\left.\left.+\gamma\frac{\Ga(-1/2-2u)\Ga(1/2-2t)}{\Ga(-2t-2u)}\right)
+3\leftrightarrow 4\right]\,\,.\labell{a718}\eeqa 
where $\z_i$ is the polarization of gauge fields. The Mandelstam variables are \beqa
s&=&-\alpha'(k_1+k_2)^2/2\,\,,\nonumber\\
t&=&-\alpha'(k_2+k_3)^2/2\,\,,\nonumber\\
u&=&-\alpha'(k_1+k_3)^2/2\,\,.\labell{mandel} \eeqa  The momenta of the gauge fields (tachyons) are $k_1,\, k_2$ ($k_3,\, k_4$). The on-shell
condition for the tachyons are $k_i^2=1/(2\alpha')$, and  the
Mandelstam variables satisfy the constraint \beqa
s+t+u&=&-1/2\labell{con3}\eeqa  The
coefficients $\alpha,\beta,\gamma$ are the non-abelian group
factors \beqa \alpha&=&\frac{1}{2}\left(\frac{}{}{\rm
Tr}(\lam_1\lam_2\lam_3\lam_4)+{\rm
Tr}(\lam_1\lam_4\lam_3\lam_2)\right)\,\,,\nonumber\\
\beta&=&\frac{1}{2}\left(\frac{}{}{\rm Tr}(\lam_1\lam_3\lam_4\lam_2)+{\rm
Tr}(\lam_1\lam_2\lam_4\lam_3)\right)\,\,,\nonumber\\
\gamma&=&\frac{1}{2}\left(\frac{}{}{\rm Tr}(\lam_1\lam_4\lam_2\lam_3)+{\rm
Tr}(\lam_1\lam_3\lam_2\lam_4)\right)\,\,.\labell{phase}\eeqa 
The standard  non-abelian kinetic terms of
the field theory  produce Feynman amplitudes that have
massless pole in  $s$-channel and  tachyonic poles in $t$- and
$u$-channels. It is shown in \cite{Garousi:2002wq} that the above amplitude reproduce the massless and tachyonic poles of the field theory if one expands the string amplitude around \beqa
s\rightarrow 0,&&  t,u\rightarrow
-1/4\eeqa
The other terms of the expansion are speculated in \cite{Garousi:2002wq, Garousi:2003pv,Garousi:2003ur} to be related to the higher derivatives of tachyon and gauge fields. In this section we would like to find these higher derivative terms to all order of $\alpha'$. To this end, we write the amplitude \reef{a718} in the 
 following form:
\beqa
A&=&4i(2\pi\alpha')T_p\left(\frac{}{}\z_1\inn\z_2 (2t')(2u')-2\alpha'\z_1\inn k_3\z_2\inn k_4(2t')-2\alpha'\z_1\inn k_4\z_2\inn k_4(2u')\right)\labell{a7188}\\
&&\times\left(\alpha
\frac{\Ga(2t'+2u')\Ga(-2t')}{\Ga(1+2u')}+\beta
\frac{\Ga(2t'+2u')\Ga(-2u')}{\Ga(1+2t')}+\gamma
\frac{\Ga(-2u')\Ga(-2t')}{\Ga(1-2t'-2u')}\right)\nonumber\eeqa
where $t'=t+1/4=-\alpha'k_2\inn k_3$ and $u'=u+1/4=-\alpha' k_1\inn k_3$. The amplitude must be expanded around  
\beqa
t',\,u'&\rightarrow &0
\eeqa 
which is the momentum expansion. Using the Maple, one can expand  the amplitude around the above point, \ie
\beqa
A&=&4i(2\pi\alpha')T_p\left(\frac{}{}\z_1\inn\z_2 (2t')(2u')-2\alpha'\z_1\inn k_3\z_2\inn k_4(2t')-2\alpha'\z_1\inn k_4\z_2\inn k_3(2u')\right)\labell{expand1}\\
&&\times\left(\frac{\alpha u'+\beta t'+\gamma s}{4t'u' s}+\sum_{n,m=0}^{\infty}\left[a_{n,m}(\alpha u'^nt'^m+\beta t'^nu'^m)+b_{n,m}\gamma(u'^nt'^m +t'^n u'^m)\right]\right)\nonumber\eeqa
where $b_{n,m}$ is symmetric. Some of the coefficients $a_{n,m}$ and $b_{n,m}$ are
\beqa
&&a_{0,0}=-\frac{\pi^2}{6},\,b_{0,0}=-\frac{\pi^2}{12}\\
&&a_{1,0}=2\z(3),\,a_{0,1}=0,\,b_{0,1}=b_{1,0}=-\z(3)\nonumber\\
&&a_{1,1}=a_{0,2}=-7\pi^4/90,\,a_{2,0}=-4\pi^4/90,\,b_{1,1}=-\pi^4/180,\,b_{0,2}=b_{2,0}=-\pi^4/45\nonumber\\
&&a_{1,2}=a_{2,1}=8\z(5)+4\pi^2\z(3)/3,\,a_{0,3}=0,\,a_{3,0}=8\z(5),\nonumber\\
&&\qquad\qquad\qquad\qquad\qquad b_{0,3}=-4\z(5),\,b_{1,2}=-8\z(5)+2\pi^2\z(3)/3\nonumber\eeqa
It has been  shown in \cite{Garousi:2002wq, Garousi:2003pv,Garousi:2003ur} that the poles in the above expansion are reproduced by the non-abelian kinetic terms, and  the contact terms with coefficients $a_{0,0}$ and $b_{0,0}$ are  also  reproduced by the following terms:
 \beqa T_p(\pi\alpha')^3{\rm
STr} \left(
\frac{}{}m^2\cT^2F_{\mu\nu}F^{\nu\mu}+\frac{}{}D^{\alpha}\cT D_{\alpha}\cT F_{\mu\nu}F^{\nu\mu}-
4F^{\mu\alpha}F_{\alpha\beta}D^{\beta}\cT D_{\mu}\cT
\right)\labell{dbicoupling} \eeqa 
where the covariant
derivative is $D_a\cT=\prt_a\cT-i[A_a,\cT]$, and STr is the symmetrised
trace prescription. They are the two tachyons and two gauge fields couplings of the nan-abelian tachyon DBI action \reef{nonab}. Writing the symmetric trace in term of ordinary trace, one can write the above couplings as
\beqa
-T_p(\pi\alpha')(\alpha')^2(\cL_{1}^{00}+\cL_{2}^{00}+\cL_{3}^{00}+\cL^{00}_{4})\labell{lagrang00}\eeqa
where
\beqa
\cL_1^{00}&=&-\frac{\pi^2}{3}m^2
\Tr\left(2\cT^2F_{\mu\nu}{F}^{\nu\mu}+ \cT F_{\mu\nu}\cT F^{\nu\mu}\right)\nonumber\\
\cL_2^{00}&=&-\frac{\pi^2}{3}\Tr\left(2D^{\alpha}\cT D_{\alpha}\cT F_{\mu\nu}{F}^{\nu\mu}+D^{\alpha} \cT F_{\mu\nu}D_{\alpha}\cT F^{\nu\mu}\right)\nonumber\\
\cL_3^{00}&=&\frac{2\pi^2}{3}\Tr
\left(2D^{\beta}\cT D_{\mu}\cT {F}^{\mu\alpha}F_{\alpha\beta}+D^{\beta}\cT{F}^{\mu\alpha}D_{\mu}\cT F_{\alpha\beta}\right)\nonumber\\
\cL_4^{00}&=&\frac{2\pi^2}{3}\Tr\left(2D^{\beta}\cT D_{\mu}\cT F_{\alpha\beta}{F}^{\mu\alpha}
+D^{\beta}\cT F_{\alpha\beta}D_{\mu}\cT {F}^{\mu\alpha}
\right)\nonumber
\eeqa
The vertex of two on-shell tachyons and two on-shell gauge fields of the above couplings  is in fact the kinematic factor in equation \reef{expand1} multiplies by $-\pi^2(\alpha+\beta+\gamma)/6$. At this order the Chan-Paton factors appear in symmetric form, \ie $\alpha+\beta+\gamma$. This is the reason that   the symmetric trace  appears in the field theory couplings \reef{dbicoupling} at this order. The Chan-Paton factors however does not appear in symmetric form in  any other  order. So one expects to have no symmetric trace in the higher order terms. 

Our strategy for finding the higher derivatives extension of the above couplings is as follows. Since the vertex of the above terms appear as coefficient of all higher order terms in \reef{expand1}, one may find the higher derivative couplings by applying   appropriate derivatives     on the above couplings.  The coefficient of each term in the above couplings is set by $a_{0,0}$ and $b_{0,0}$. In the higher derivative extensions  one should replace them by  $a_{n,m}$ and $b_{n,m}$. Let us focus on $\cL_1^{00}$ terms which appear in the above couplings as
\beqa
\cL_1^{00}&=&m^2\Tr\left(4a_{0,0}\cT^2F_{\mu\nu}{F}^{\nu\mu}+4b_{0,0} \cT F_{\mu\nu}\cT F^{\nu\mu}\right)\eeqa
Extension of $a_{0,0}$  to $a_{1,0}$  is the following:
\beqa
&&m^2\left(\alpha'\right)\Tr\left(\frac{}{}2a_{1,0}[D_a\cT \cT D^aF_{\mu\nu}{F}^{\nu\mu}+\cT D_a\cT F_{\mu\nu}D^a{F}^{\nu\mu}]\right)\nonumber\\
&&=m^2\left(\alpha'\right)\Tr\left(\frac{}{}2a_{1,0}[D_a\cT \cT D^aF_{\mu\nu}{F}^{\nu\mu}+h.c.]\right)\eeqa
where $D_aF_{\mu\nu}=\prt_a F_{\mu\nu}-i[A_a,F_{\mu\nu}]$. Note that the above Lagrangian  is hermitian. Further extension to $a_{1,1}$ is 
\beqa
m^2\left(\alpha'\right)^2\Tr\left(\frac{}{}a_{1,1}[D_aD_b\cT \cT D^aF_{\mu\nu}D^b{F}^{\nu\mu}+ D_aD_bF_{\mu\nu}{F}^{\nu\mu}D^a\cT D^b\cT+h.c.]\right)\eeqa
 Now, it is not difficult to extend the above couplings    to $a_{n,m}$  case, \ie
\beqa
&&m^2\left(\alpha'\right)^{n+m}a_{n,m}
\Tr\left(\frac{}{}\cD_{nm}(\cT \cT F_{\mu\nu}F^{\nu\mu})+\cD_{nm}(F_{\mu\nu} F^{\nu\mu} \cT  \cT)
+h.c.\right)\labell{expand2}\eeqa
where the higher derivative operator $\cD_{nm}$ is defined as
\beqa
\cD_{nm}(EFGH)&\equiv&D_{b_1}\cdots D_{b_m}D_{a_1}\cdots D_{a_n}E  F D^{a_1}\cdots D^{a_n}GD^{b_1}\cdots D^{b_m}H\eeqa
Similarly, the extension of $b_{0,0}$  to $b_{1,0}$  is the following:
\beqa
&&m^2\left(\alpha'\right)\Tr\left(\frac{}{}2b_{1,0}[D_a\cT D^a F_{\mu\nu}\cT{F}^{\nu\mu}+h.c]\right) \nonumber\eeqa
Further extension to $b_{1,1}$ is 
\beqa
m^2\left(\alpha'\right)^2\Tr\left(\frac{}{}b_{1,1}[D_aD_b\cT D^a F_{\mu\nu}\cT D^b{F}^{\nu\mu}+ D_aD_bF_{\mu\nu}D^a\cT {F}^{\nu\mu} D^b\cT+h.c.]\right)\eeqa
and  extension      to $b_{n,m}$  case is
\beqa
&&m^2\left(\alpha'\right)^{n+m}a_{n,m}
\Tr\left(\frac{}{}\cD'_{nm}(\cT  F_{\mu\nu}\cT F^{\nu\mu})+\cD'_{nm}(F_{\mu\nu} \cT F^{\nu\mu} \cT  )
+h.c.\right)\labell{expand20}\eeqa
where the higher derivative operator $\cD'_{nm}$ is defined as
\beqa
\cD'_{nm}(EFGH)&\equiv&D_{b_1}\cdots D_{b_m}D_{a_1}\cdots D_{a_n}E   D^{a_1}\cdots D^{a_n}F G D^{b_1}\cdots D^{b_m}H\eeqa
One can repeat similar steps for all other terms in \reef{lagrang00}. Hence, the proposal for the couplings between two tachyons and two  field strengths on the world volume of $N$ non-BPS D-branes, to all order of $\alpha'$, is the following:
\beqa
\cL_{}&=&-T_p(\pi\alpha')(\alpha')^{2+n+m}\sum_{n,m=0}^{\infty}(\cL_{1}^{nm}+\cL_{2}^{nm}+\cL_{3}^{nm}+\cL^{nm}_{4})\labell{lagrang}\eeqa
where
\beqa
\cL_1^{nm}&=&m^2
\Tr\left(\frac{}{}a_{n,m}[\cD_{nm}(\cT^2F_{\mu\nu}{F}^{\nu\mu})+ \cD_{nm}(F_{\mu\nu}{F}^{\nu\mu}\cT^2)]\right.\nonumber\\
&&\left.+\frac{}{}b_{n,m}[\cD'_{nm}(\cT F_{\mu\nu}\cT F^{\nu\mu})+\cD'_{nm}( F_{\mu\nu}\cT F^{\nu\mu}\cT)]+h.c.\right)\nonumber\\
\cL_2^{nm}&=&\Tr\left(\frac{}{}a_{n,m}[\cD_{nm}(D^{\alpha}\cT D_{\alpha}\cT F_{\mu\nu}{F}^{\nu\mu})+\cD_{nm}( F_{\mu\nu}{F}^{\nu\mu}D^{\alpha}\cT D_{\alpha}\cT)]\right.\nonumber\\
&&\left.+\frac{}{}b_{n,m}[\cD'_{nm}(D^{\alpha} \cT F_{\mu\nu}D_{\alpha}\cT F^{\nu\mu})+\cD'_{nm}( F_{\mu\nu}D_{\alpha}\cT F^{\nu\mu}D^{\alpha} \cT)]+h.c.\right)\nonumber\\
\cL_3^{nm}&=&-2\Tr
\left(\frac{}{}a_{n,m}[\cD_{nm}(D^{\beta}\cT D_{\mu}\cT {F}^{\mu\alpha}F_{\alpha\beta})+\cD_{nm}( {F}^{\mu\alpha}F_{\alpha\beta}D^{\beta}\cT D_{\mu}\cT)]\right.\nonumber\\
&&\left.+\frac{}{}b_{n,m}[\cD'_{nm}(D^{\beta}\cT{F}^{\mu\alpha}D_{\mu}\cT F_{\alpha\beta})+\cD'_{nm}({F}^{\mu\alpha}D_{\mu}\cT F_{\alpha\beta}D^{\beta}\cT)]+h.c.\right)\nonumber\\
\cL_4^{nm}&=&-2\Tr\left(\frac{}{}a_{n,m}[\cD_{nm}(D^{\beta}\cT D_{\mu}\cT F_{\alpha\beta}{F}^{\mu\alpha})
+\cD_{nm}( F_{\alpha\beta}{F}^{\mu\alpha}D^{\beta}\cT D_{\mu}\cT)]\right.\nonumber\\
&&\left.+\frac{}{}b_{n,m}[\cD'_{nm}(D^{\beta}\cT F_{\alpha\beta}D_{\mu}\cT {F}^{\mu\alpha})+\cD'_{nm}( F_{\alpha\beta}D_{\mu}\cT {F}^{\mu\alpha}D^{\beta}\cT)]+h.c.
\right)\nonumber
\eeqa
  If one calculates the coupling of two on-shell tachyons and two gauge fields from \reef{lagrang}, one will find the contact  terms in the amplitude \reef{expand1}. 
  
When  the covariant derivative of the field strength and the second covariant derivative of tachyon are  zero, the Lagrangian \reef{lagrang}  reduces to the couplings \reef{dbicoupling} which are the two-tachyons-two-gauge field strengths couplings of the non-abelian tachyon DBI action. This may indicates that  the non-abelian tachyon DBI action is   the effective action of the non-BPS D-branes when fields  vary slowly. 

The couplings in \reef{lagrang} have  on-shell ambiguity, \ie $\cT\sim 2\alpha'\prt_a\prt^a\cT$. This on-shell ambiguity has no effect on the massless and the simple tachyon poles of the S-matrix elements. In the massless pole it is obvious because the tachyons are on-shell. In the tachyon pole $\cT$ appears as $1/(k^2-1/2\alpha')$, whereas  $2\alpha'\prt_a\prt^a\cT$ appears as
$2\alpha'k^2/(k^2-1/2\alpha')$, however, one can write it as
\beqa
\frac{2\alpha'k^2}{k^2-1/2\alpha'}&=&\frac{1}{k^2-1/2\alpha'}+2\alpha'\eeqa
hence in the tachyon pole  $\cT$ and $2\alpha'\prt_a\prt^a\cT$ have identical effect. However, the deference  is an extra contact term. By studying a S-matrix element in which the couplings \reef{lagrang} appear in tachyon poles as well as the contact terms, one way fix the on-shell ambiguity of \reef{lagrang}.

\subsection{Four massless scalars couplings}

It has been shown in \cite{ Garousi:2003pv,Garousi:2003ur} that the S-matrix element of four massless transverse scalars and the S-matrix element of four tachyons can be written in a universal form. So, one may expect that the higher derivative couplings of four tachyons should be similar to the higher derivative couplings of four scalar fields. In fact the tachyon and the scalar fields appear in similar form in the tachyon DBI action. The only difference is that there is a  potential for the tachyon, \eg  $e^{\pi\alpha'm^2T^2}$ where $m^2$ is the mass of the tachyon. Therefore, to find the higher derivative couplings of four tachyons, we first find the higher derivative couplings of four scalar fields and then inspired by them we will find the tachyon couplings.

The S-matrix element of four massless transverse scalar vertex operators
in the supersting theory   is given by
$A=A_s+A_t+A_u$ where \beqa A
_s&\!\!\!=\!\!\!&-4iT_p\z_1\inn\z_2\z_3\inn\z_4\left(\alpha\frac{\Ga(-2s)
\Ga(1-2t)}{\Ga(-2s-2t)}+
\beta\frac{\Ga(-2s)\Ga(1-2u)}{\Ga(-2s-2u)}-
\gamma\frac{\Ga(1-2t)\Ga(1-2u)}{\Ga(1-2t-2u)}\right)\nonumber\\
A
_u&\!\!\!=\!\!\!&-4iT_p\z_1\inn\z_3\z_2\inn\z_4\left(-\alpha\frac{\Ga(1-2s)
\Ga(1-2t)}{\Ga(1-2s-2t)}+
\beta\frac{\Ga(-2u)\Ga(1-2s)}{\Ga(-2u-2s)}+
\gamma\frac{\Ga(-2u)\Ga(1-2t)}{\Ga(-2u-2t)}\right)\nonumber\\
A
_t&\!\!\!=\!\!\!&-4iT_p\z_1\inn\z_4\z_2\inn\z_3\left(\alpha\frac{\Ga(-2t)
\Ga(1-2s)}{\Ga(-2t-2s)}-
\beta\frac{\Ga(1-2s)\Ga(1-2u)}{\Ga(1-2s-2u)}+\gamma
\frac{\Ga(-2t)\Ga(1-2u)}{\Ga(-2t-2u)}\right)\nonumber\labell{a7}\eeqa
where $\z$'s are the scalars polarization.  The on-shell condition
for the scalars are $k_i^2=0$, and the Mandelstam variables
constrain to the relation \beqa s+t+u&=&0\labell{con2}\eeqa  In
this case the massless poles of the Feynman amplitude resulting from the non-abelian kinetic term of the scalars can be produced by the above amplitude expanded at low energy, \ie  $s,t,u\rightarrow 0$. To find the four scalars couplings to all order of $\alpha'$, we repeat the steps in the previous section, so rewrite the amplitudes as
\beqa A
_s&=&16iT_p\z_1\inn\z_2\z_3\inn\z_4\times\nonumber\\
&& tu\left(\alpha\frac{\Ga(2t+2u) \Ga(-2t)}{\Ga(1+2u)}+
\beta\frac{\Ga(2t+2u)\Ga(-2u)}{\Ga(1+2t)}+
\gamma\frac{\Ga(-2t)\Ga(-2u)}{\Ga(1-2t-2u)}\right)\nonumber\\
A_u&=&16iT_p \z_1\inn\z_3\z_2\inn\z_4\times\nonumber\\
&&ts\left(\alpha\frac{\Ga(-2s) \Ga(-2t)}{\Ga(1-2s-2t)}+
\beta\frac{\Ga(2t+2s)\Ga(-2s)}{\Ga(1+2t)}+
\gamma\frac{\Ga(2t+2s)\Ga(-2t)}{\Ga(1+2s)}\right)\nonumber\\
A _t&=&16iT_p \z_1\inn\z_4\z_2\inn\z_3\times\nonumber\\
&&us\left(\alpha\frac{\Ga(2s+2u) \Ga(-2s)}{\Ga(1+2u)}+
\beta\frac{\Ga(-2u)\Ga(-2s)}{\Ga(1-2s-2u)}+
\gamma\frac{\Ga(2u+2s)\Ga(-2u)}{\Ga(1+2s)}\right)\nonumber\eeqa 
Using the Maple, one can expand  the amplitude around $s,t,u\rightarrow 0$, \ie
\beqa
A_s&=&16iT_p\z_1\inn\z_2\z_3\inn\z_4tu\times\labell{**}\\
&& \left(\frac{\alpha u+\beta t+\gamma s}{4tu s}+\sum_{n,m=0}^{\infty}\left[a_{n,m}(\alpha u^nt^m+\beta t^nu^m)+b_{n,m}\gamma(u^nt^m+t^nu^m)\right]\right)\nonumber\\
A_u&=&16iT_p\z_1\inn\z_3\z_2\inn\z_4ts\times\nonumber\\
&&\left(\frac{\gamma s+\beta t+\alpha u}{4ts u}+\sum_{n,m=0}^{\infty}\left[a_{n,m}(\gamma s^nt^m+\beta t^ns^m)+b_{n,m}\alpha(s^nt^m+t^ns^m)\right]\right)\nonumber\\
A_t&=&16iT_p\z_1\inn\z_4\z_2\inn\z_3 su\times\nonumber\\
&&\left(\frac{\alpha u+\gamma s+\beta t}{4su t}+\sum_{n,m=0}^{\infty}\left[a_{n,m}(\alpha u^ns^m+\gamma s^nu^m)+b_{n,m}\beta(u^ns^m+s^nu^m)\right]\right)\nonumber\eeqa
 On can also write the lase term in each amplitude in another form, \eg the last term in the first line can be written also as $\sum_{n,m=0}^{\infty}(tu)^m(s)^n$. They produce  different four scalars couplings. Up to total derivative terms,  the differences are in the couplings which involve $\prt_a\prt^a\phi^i$.  They have no effect on the simple massless poles of S-matrix elements because   canceling $k^2$ with the massless propagator one finds a contact term.
   
The massless poles in \reef{**} are reproduced by the non-abelian kinetic terms of the scalar field, and  the contact terms with coefficients $a_{0,0}$ and $b_{0,0}$ are  also  reproduced by the following terms:
 \beqa
&&- T_p{\rm STr}
\left(-\frac{1}{4}D_a\phi^iD_b\phi_iD^b\phi^jD^a\phi_j+\frac{1}{8}
(D_a\phi^i D^a\phi_i)^2\right)\,\, \labell{a011}\eeqa 
Writing the symmetric trace in term of ordinary trace, one can write it as
\beqa
T_p(\cL_{5}^{00}+\cL_{6}^{00}+\cL_{7}^{00})\labell{lagrang001}\eeqa
where
\beqa
\cL_5^{00}&=&-\frac{1}{8\pi^2}
\Tr\left(4a_{0,0}D_{\alpha}\phi^iD_{\beta}\phi_iD^{\beta}\phi^jD^{\alpha}\phi_j+ 4b_{0,0} D_{\alpha}\phi^iD^{\beta}\phi^jD_{\beta}\phi_iD^{\alpha}\phi_j\right)\nonumber\\
\cL_6^{00}&=&-\frac{1}{8\pi^2}
\Tr\left(4a_{0,0}D_{\alpha}\phi^iD_{\beta}\phi_iD^{\alpha}\phi^jD^{\beta}\phi_j+ 4b_{0,0} D_{\alpha}\phi^iD^{\alpha}\phi^jD_{\beta}\phi_iD^{\beta}\phi_j\right)\nonumber\\
\cL_{7}^{00}&=&\frac{1}{8\pi^2}
\Tr\left(4a_{0,0}D_{\alpha}\phi^iD^{\alpha}\phi_iD_{\beta}\phi^jD^{\beta}\phi_j+ 4b_{0,0} D_{\alpha}\phi^iD_{\beta}\phi^jD^{\alpha}\phi_iD^{\beta}\phi_j\right)\nonumber
\eeqa
To check the consistency of the above couplings with the $(a_{0,0},b_{0,0})$ order contact terms of the string theory amplitude \reef{**}, one uses the relation $2k_1\inn k_2\, k_1\inn k_3=-(k_1\inn k_2)(k_3\inn k_4)-(k_1\inn k_3)(k_2\inn k_4)+(k_2\inn k_3)(k_1\inn k_4)$. Now one can extend it easily to the higher derivative terms as
\beqa
\frac{1}{4\pi^2}T_p\left(\alpha'\right)^{n+m}\sum_{m,n=0}^{\infty}(\cL_{5}^{nm}+\cL_{6}^{nm}+\cL_{7}^{nm})\labell{lagrang1}\eeqa
where
\beqa
&&\cL_{5}^{nm}=-
\Tr\left(\frac{}{}a_{n,m}\cD_{nm}[D_{\alpha}\phi^i D_{\beta}\phi_i D^{\beta}\phi^j D^{\alpha}\phi_j]+\frac{}{} b_{n,m}\cD'_{nm}[D_{\alpha}\phi^i D^{\beta}\phi^j  D_{\beta}\phi_i  D^{\alpha}\phi_j ]+h.c.\frac{}{}\right)\nonumber\\
&&\cL_{6}^{nm}=-\Tr\left(\frac{}{}a_{n,m}\cD_{nm}[D_{\alpha}\phi^i D_{\beta}\phi_i D^{\alpha}\phi^j D^{\beta}\phi_j]+\frac{}{}b_{n,m}\cD'_{nm}[D_{\beta}\phi^i D^{\beta}\phi^j  D_{\alpha}\phi_i  D^{\alpha}\phi_j  ]+h.c.\frac{}{}\right)\nonumber\\
&&\cL_{7}^{nm}=\Tr\left(\frac{}{}a_{n,m}\cD_{nm}[D_{\alpha}\phi^i D^{\alpha}\phi_i D_{\beta}\phi^j D^{\beta}\phi_j]+\frac{}{}b_{n,m}\cD'_{nm}[D_{\alpha}\phi^i D_{\beta}\phi^j  D^{\alpha}\phi_i  D^{\beta}\phi_j ]+h.c\frac{}{}\right)\nonumber\eeqa
It is not difficult to check that the infinite tower of four scalar couplings \reef{lagrang1} produce the string theory S-matrix element \reef{**}. The above couplings can be extended  to the coupling of four gauge fields by using T-duality. Up to total derivative terms and the terms like   $FFF\prt_a\prt^a F$, which is zero on-shell, one can show that the $FFD F DF$ terms are those appear in the literature. We now turn to the couplings of four tachyons.

\subsection{Four tachyons couplings}

The S-matrix element of four open string tachyon vertex operators
in the supersting theory   is given by \cite{mgjs,jp} \beqa A&=&-12iT_p\left(\alpha\frac{\Ga(-2t)\Ga(-2s)}{\Ga(-1-2t-2s)}+
\beta\frac{\Ga(-2s)\Ga(-2u)}{\Ga(-1-2s-2u)}+
\gamma\frac{\Ga(-2t)\Ga(-2u)}{\Ga(-1-2t-2u)}\right)\labell{a1} \eeqa
where the Mandelstam variables are those defined in \reef{mandel}
 and   satisfy the constraint \beqa
s+t+u&=&-1\labell{con1}\,\,. \eeqa
The standard non-abelian kinetic term in field theory produces
massless poles in $s$-, $t$-,  $u$-channels. However, the
constraint \reef{con1} does not allow us to sent all $s,t,u$ to
zero at the same time to produce
massless poles. It is shown in
\cite{Garousi:2002wq,Garousi:2003pv,Garousi:2003ur} that in order to produce the massless poles, one
should first arrange the amplitude in a specific form, \ie
 one should
write $A=A_s+A_t+A_u$ where \beqa A
_s&=&-4iT_p\left(\alpha\frac{\Ga(-2s) \Ga(-2t)}{\Ga(-1-2s-2t)}+
\beta\frac{\Ga(-2s)\Ga(-2u)}{\Ga(-1-2s-2u)}-
\gamma\frac{\Ga(-2t)\Ga(-2u)}{\Ga(-1-2t-2u)}\right)\nonumber\\
A _u&=&-4iT_p\left(-\alpha\frac{\Ga(-2s)
\Ga(-2t)}{\Ga(-1-2s-2t)}+
\beta\frac{\Ga(-2u)\Ga(-2s)}{\Ga(-1-2u-2s)}+
\gamma\frac{\Ga(-2u)\Ga(-2t)}{\Ga(-1-2u-2t)}\right)\nonumber\\
A_t&=&-4iT_p\left(\alpha\frac{\Ga(-2t)
\Ga(-2s)}{\Ga(-1-2t-2s)}-
\beta\frac{\Ga(-2s)\Ga(-2u)}{\Ga(-1-2s-2u)}+\gamma
\frac{\Ga(-2t)\Ga(-2u)}{\Ga(-1-2t-2u)}\right)\labell{a717}\eeqa Then one should send 
 \beqa s-{\rm
channel}:&&\lim_{s\rightarrow 0\,,t,u\rightarrow -1/2}A_s\nonumber\\
t-{\rm channel}:&&\lim_{t\rightarrow 0\,,s,u\rightarrow
-1/2}A_t\nonumber\\
u-{\rm channel}:&&\lim_{u\rightarrow 0\,,s,t\rightarrow
-1/2}A_u\labell{lim1}\eeqa 
These limits  are consistent with the constraint \reef{con1}. 

The leading term of this expansion produces the the massless  poles of the higher derivative  theory, and the other terms are speculated  to be related to the higher derivatives of the tachyon \cite{ Garousi:2002wq,Garousi:2003pv,Garousi:2003ur}. In this section we would like to find these higher derivative terms up to on-shell ambiguity. So we write the amplitude in the 
 following form:
\beqa A
_s&\!\!\!=\!\!\!&16iT_pt'u'\left(\alpha\frac{\Ga(2t'+2u') \Ga(-2t')}{\Ga(1+2u')}+
\beta\frac{\Ga(2t'+2u')\Ga(-2u')}{\Ga(1+2t')}+
\gamma\frac{\Ga(-2t')\Ga(-2u')}{\Ga(1-2t'-2u')}\right)\nonumber\\
A_u&\!\!\!=\!\!\!&16iT_pt's'\left(\alpha\frac{\Ga(-2s') \Ga(-2t')}{\Ga(1-2s'-2t')}+
\beta\frac{\Ga(2t'+2s')\Ga(-2s')}{\Ga(1+2t')}+
\gamma\frac{\Ga(2t'+2s')\Ga(-2t')}{\Ga(1+2s')}\right)\label{a719}\\
A_t&\!\!\!=\!\!\!&16iT_pu's'\left(\alpha\frac{\Ga(2s'+2u') \Ga(-2s')}{\Ga(1+2u')}+
\beta\frac{\Ga(-2u')\Ga(-2s')}{\Ga(1-2s'-2u')}+
\gamma\frac{\Ga(2u'+2s')\Ga(-2u')}{\Ga(1+2s')}\right)\nonumber\eeqa 
where $s'=s+1/2=-\alpha'k_1\inn k_2$, $t'=t+1/2=-\alpha'k_2\inn k_3$ and $u'=u+1/2=-\alpha'k_1\inn k_3$. Now the field theory corresponds to expanding the above amplitude around   \beqa
s',\,t',\, u'&\rightarrow &0\eeqa
which is the momentum expansion. Note that $s+t'+u'=0$, $u+s'+t'=0$ and $t+s'+u'=0$. Like the scalar case, one should expand   the amplitude around the above point, \ie
\beqa
A_s&=&16iT_pt'u'\times\nonumber\\
&&\left(\frac{\alpha u'+\beta t'+\gamma s}{4t'u' s}+\sum_{n,m=0}^{\infty}\left[a_{n,m}(\alpha u'^nt'^m+\beta t'^nu'^m)+b_{n,m}\gamma(u'^nt'^m+ t'^nu'^m)\right]\right)\nonumber\\
A_u&=&16iT_pt's'\times\nonumber\\
&&\left(\frac{\gamma s'+\beta t'+\alpha u}{4t's' u}+\sum_{n,m=0}^{\infty}\left[a_{n,m}(\gamma s'^nt'^m+\beta t'^ns'^m)+b_{n,m}\alpha(s'^nt'^m+ t'^ns'^m)\right]\right)\nonumber\\
A_t&=&16iT_ps'u'\times\nonumber\\
&&\left(\frac{\alpha u'+\gamma s'+\beta t}{4s'u' t}+\sum_{n,m=0}^{\infty}\left[a_{n,m}(\alpha u'^ns'^m+\gamma s'^nu'^m)+b_{n,m}\beta(u'^ns'^m+ s'^nu'^m)\right]\right)\nonumber\eeqa
The poles in the above expansion are reproduced by the non-abelian kinetic terms,  and  the contact terms with coefficients $a_{0,0}$ and $b_{0,0}$ are    reproduced by the following terms \cite{Garousi:2002wq,Garousi:2003pv,Garousi:2003ur}:
 \beqa
&&-(2\pi\alpha')^2 T_p{\rm STr}
\left(\frac{m^4}{8}\cT^4+\frac{m^2}{4}\cT^2D_a\cT D^a\cT-\frac{1}{8}
(D_a\cT D^a\cT)^2\right)\,\, \labell{a01}\eeqa 
which are the four tachyons coupling of the non-abelian tachyon DBI action. To check this explicitly, one needs the on-shell  relation $2k_1\inn k_2\, k_1\inn k_3=m^4-m^2(k_2\inn k_3+k_1\inn k_4)-(k_1\inn k_2)(k_3\inn k_4)-(k_1\inn k_3)(k_2\inn k_4)+(k_2\inn k_3)(k_1\inn k_4)$. Writing the symmetric trace in term of ordinary trace, one can write \reef{a01}  as
\beqa
T_p(\alpha')^2(\cL_{8}^{00}+\cL_{9}^{00}+\cL_{10}^{00}+\cL_{11}^{00}+\cL_{12}^{00})\labell{lagrang0011}\eeqa
where
\beqa
\cL_8^{00}&=&2m^4
\Tr\left(a_{0,0}\cT^4+ b_{0,0}\cT^4 \right)\nonumber\\
\cL_9^{00}&=&4m^2\Tr\left(a_{0,0}\cT^2D^{\alpha}\cT D_{\alpha}\cT +b_{0,0}\cT D^{\alpha} \cT \cT D_{\alpha}\cT \right)\nonumber\\
\cL_{10}^{00}&=&-2
\Tr\left(a_{0,0}D_{\alpha}\cT D_{\beta}\cT D^{\beta}\cT D^{\alpha}\cT+ b_{0,0} D_{\alpha}\cT D^{\beta}\cT D_{\beta}\cT  D^{\alpha}\cT \right)\nonumber\\
\cL_{11}^{00}&=&-2
\Tr\left(a_{0,0}D_{\alpha}\cT D_{\beta}\cT D^{\alpha}\cT D^{\beta}\cT+ b_{0,0} D_{\alpha}\cT D^{\alpha}\cT 
     D_{\beta}\cT D^{\beta}\cT \right)\nonumber\\
\cL_{12}^{00}&=&2
\Tr\left(a_{0,0}D_{\alpha}\cT D^{\alpha}\cT D_{\beta}\cT D^{\beta}\cT+ b_{0,0} D_{\alpha}\cT D_{\beta}\cT  D^{\alpha}\cT D^{\beta}\cT\right)\labell{L012}
\eeqa
Note that the last three lines  add up to 
\beqa
\frac{\pi^2}{6}\Tr
\left(2D^{\alpha}\cT D_{\alpha}\cT D^{\beta}\cT D_{\beta}\cT +D^{\alpha}\cT D^{\beta}\cT D_{\alpha}\cT D_{\beta}\cT \right)\nonumber
\eeqa
However, we have written them in the form \reef{L012} to extend them easily to the higher derivative terms using the fact that the higher derivative extension of \reef{lagrang001} is \reef{lagrang1}. The higher derivative extension of $\cL_9^{00}$ is like the higher derivative extension of $\cL_2^{00}$ which appears in \reef{lagrang}. Therefore, the 
 higher derivative extension of \reef{lagrang0011} is the following:
\beqa
T_p(\alpha')^{2+n+m}\sum_{m,n=0}^{\infty}(\cL_{8}^{nm}+\cL_{9}^{nm}+\cL_{10}^{nm}+\cL_{11}^{nm}+\cL_{12}^{nm})\labell{lagrang11}\eeqa
where
\beqa
\cL_{8}^{nm}&\!\!\!\!\!\!=\!\!\!\!\!\!&m^4\Tr\left(\frac{}{}a_{n,m}\cD_{nm}[\cT\cT\cT\cT]+ b_{n,m}\cD'_{nm}[\cT\cT\cT\cT]+h.c.\frac{}{}\right)\nonumber\\
\cL_{9}^{nm}&\!\!\!\!\!\!=\!\!\!\!\!\!&m^2\Tr\left(\frac{}{}a_{n,m}[\cD_{nm}(\cT \cT D^{\alpha}\cT D_{\alpha}\cT)+\cD_{nm}( D^{\alpha}\cT D_{\alpha}\cT\cT \cT)]\right.\nonumber\\
&&\left.+b_{n,m}[\cD'_{nm}(\cT D^{\alpha}\cT  \cT  D_{\alpha}\cT)+\cD'_{nm}( D^{\alpha}\cT  \cT  D_{\alpha}\cT \cT)] +h.c.\frac{}{}\right)\nonumber\\
\cL_{10}^{nm}&\!\!\!\!\!\!=\!\!\!\!\!\!&-\Tr\left(\frac{}{}a_{n,m}\cD_{nm}[D_{\alpha}\cT D_{\beta}\cT D^{\beta}\cT D^{\alpha}\cT]+b_{n,m}\cD'_{nm}[D_{\alpha}\cT D^{\beta}\cT D_{\beta}\cT  D^{\alpha}\cT]+h.c.\frac{}{}\right)\nonumber\\
\cL_{11}^{nm}&\!\!\!\!\!\!=\!\!\!\!\!\!&-\Tr\left(\frac{}{}a_{n,m}\cD_{nm}[D_{\alpha}\cT D_{\beta}\cT D^{\alpha}\cT D^{\beta}\cT] +b_{n,m}\cD'_{nm}[D_{\beta}\cT D^{\beta}\cT  D_{\alpha}\cT D^{\alpha}\cT]+h.c. \frac{}{}\right)\nonumber\\
\cL_{12}^{nm}&\!\!\!\!\!\!=\!\!\!\!\!\!&\Tr\left(\frac{}{}a_{n,m}\cD_{nm}[D_{\alpha}\cT D^{\alpha}\cT D_{\beta}\cT D^{\beta} \cT] +b_{n,m}\cD'_{nm}[D_{\alpha}\cT D_{\beta}\cT D^{\alpha}\cT  D^{\beta}\cT]+h.c. \frac{}{}\right)\nonumber\eeqa
It is not difficult to check that the infinite tower of four tachyons couplings \reef{lagrang11} produce the string theory S-matrix element. It is interesting to note that  when the second derivative  of tachyon is zero, the Lagrangian \reef{lagrang11}  reduces to DBI couplings \reef{a01} for abelian tachyon. For nonabelian case, it does not reduces to the tachyon couplings of the non-abelian DBI action.  However, as we mentioned before
the couplings \reef{lagrang11}  have on-shell ambiguity, \ie $\cT\sim 2\alpha'\prt_a\prt^a\cT$.  As long as the on-shell ambiguity is not  fixed, one can not use the tachyon couplings \reef{lagrang11} to find the effective tachyon couplings. It has been check in \cite{BitaghsirFadafan:2006cj} that the couplings \reef{a01} have no on-shell ambiguity, \ie these couplings appear in the tachyon pole and contact terms of the S-matrix element of four tachyons and one gauge field at $\z(2)$ order.   It would be interesting to compare the infinite tachyon poles and contact terms of the S-matrix element found in \cite{BitaghsirFadafan:2006cj} with the tachyon pole and the contact terms of the scattering amplitude of four tachyons and one gauge field using the infinite tower of tachyon couplings in \reef{lagrang11}.  This calculation may fix the on-shell ambiguity of all couplings in \reef{lagrang11}.

\section{Higher derivative terms of brane-anti-brane}\label{scatt}

Having found the couplings in the world-volume theory of $N$ non-BPS D-brane, we can now find the corresponding couplings in the world-volume theory of brane-anti-brane. It has been  proposed in \cite{Garousi:2004rd} that the  action of D-brane-anti-D-brane may be given by the projection of the  action of two non-BPS D-brane with $(-1)^{F_L}$ where $F_L$ is the spacetime left hand fermion number. According to this proposal,  the couplings in the brane-anti-brane theory  can be read from the non-BPS branes  couplings by using the   matrices  \reef{M12} for the field strength and for the tachyon.
If one replaces them   into \reef{lagrang} and performing the trace, one finds the following  couplings between two tachyons and two gauge fields: 
\beqa
\cL_{D\bD}&=&-T_p(\pi\alpha')(\alpha')^{2+n+m}\sum_{n,m=0}^{\infty}(\cL_{1D\bD}^{nm}+\cL_{2D\bD}^{nm}+\cL_{3D\bD}^{nm}+\cL^{nm}_{4D\bD})\labell{ddbarcoupling}\eeqa
where
\beqa
\cL_{1D\bD}^{nm}&=&m^2
\left(\frac{}{}a_{n,m}[\cD_{nm}(TT^*F^{(1)}_{\mu\nu}{F}^{(1)\nu\mu})+ \cD_{nm}(F^{(1)}_{\mu\nu}{F}^{(1)\nu\mu} TT^*)+c.c]\right.\nonumber\\
&&\left.+\frac{}{}b_{n,m}[\cD'_{nm}(T F^{(2)}_{\mu\nu}T^* F^{(1)\nu\mu})+\cD'_{nm}( F^{(1)}_{\mu\nu}T F^{(2)\nu\mu}T^*)+c.c.]\right)\nonumber\\
\cL_{2D\bD}^{nm}&=&\left(\frac{}{}a_{n,m}[\cD_{nm}(D^{\alpha}T D_{\alpha}T^* F^{(1)}_{\mu\nu}{F}^{(1)\nu\mu})+\cD_{nm}( F^{(1)}_{\mu\nu}{F}^{(1)\nu\mu}D^{\alpha}T D_{\alpha}T^*)+c.c]\right.\nonumber\\
&&\left.+\frac{}{}b_{n,m}[\cD'_{nm}(D^{\alpha} T F^{(2)}_{\mu\nu}D_{\alpha}T^* F^{(1)\nu\mu})+\cD'_{nm}( F^{(1)}_{\mu\nu}D_{\alpha}T F^{(2)\nu\mu}D^{\alpha} T^*)+c.c]\right)\nonumber\\
\cL_{3D\bD}^{nm}&=&-2
\left(\frac{}{}a_{n,m}[\cD_{nm}(D^{\beta}T D_{\mu}T^* {F}^{(1)\mu\alpha}F^{(1)}_{\alpha\beta})+\cD_{nm}( {F}^{(1)\mu\alpha}F^{(1)}_{\alpha\beta}D^{\beta}T D_{\mu}T^*)+c.c]\right.\nonumber\\
&&\left.+\frac{}{}b_{n,m}[\cD'_{nm}(D^{\beta}T{F}^{(2)\mu\alpha}D_{\mu}T^* F^{(1)}_{\alpha\beta})+\cD'_{nm}({F}^{(1)\mu\alpha}D_{\mu}T F^{(2)}_{\alpha\beta}D^{\beta}T^*)+c.c.]\right)\nonumber\\
\cL_{4D\bD}^{nm}&=&-2\left(\frac{}{}a_{n,m}[\cD_{nm}(D^{\beta}T D_{\mu}T^* F^{(1)}_{\alpha\beta}{F}^{(1)\mu\alpha})
+\cD_{nm}( F^{(1)}_{\alpha\beta}{F}^{(1)\mu\alpha}D^{\beta}T D_{\mu}T^*)+c.c]\right.\nonumber\\
&&\left.+\frac{}{}b_{n,m}[\cD'_{nm}(D^{\beta}T F^{(2)}_{\alpha\beta}D_{\mu}T^* {F}^{(1)\mu\alpha})+\cD'_{nm}( F^{(1)}_{\alpha\beta}D_{\mu}T {F}^{(2)\mu\alpha}D^{\beta}T^*)+c.c.]
\right)\nonumber
\eeqa
plus $F^{(1)}\rightarrow F^{(2)}$ for $a_{n,m}$ terms and $F^{(1)}\leftrightarrow F^{(2)}$ for $b_{n,m}$ terms. In above equation, the covariant derivative of field strength is ordinary derivative and  $D_{a_1}\cdots D_{a_n}T=\prt_{a_1}D_{a_2}\cdots D_{a_{n}}T-i(A^{(1)}_{a_1}-A^{(2)}_{a_1})D_{a_2}\cdots D_{a_{n}}T$. Note that there are couplings between $F^{(1)}$ and $F^{(2)}$. For the case $T=T^*$ and $F^{(1)}=F^{(2)}$, the Lagrangian \reef{ddbarcoupling} is the same as the  Lagrangian \reef{lagrang} for one non-BPS D-brane as expected. One can do similar steps to find the four tachyon couplings in the brane-anti-brane world-volume  theory. 

We have found the tachyon couplings  from the contact terms of  the momentum expansion of the string theory S-matrix elements. A nontrivial consistency check of the higher derivative  theory with string theory is that the tachyon couplings  reproduce also the infinite massless or tachyonic poles of the string theory S-matrix elements. As an example we consider the momentum expansion of the S-matrix element of one RR field $C_{p-1}$, two tachyons and one gauge field which is nonzero in the world volume theory of brane-anti-brane. This expansion has infinite contact terms and also infinite massless and tachyonic poles \cite{Garousi:2007si}. The contact terms and the tachyonic poles are reproduced by appropriate couplings of one RR field, two tachyons and one gauge field in field theory which has been found in \cite{Garousi:2007si}, \ie the higher derivative couplings in the Introduction section. The massless poles are the following:
\beqa
&&i\mu_p(\alpha')^2\frac{ \eps^{a_{0}\cdots a_{p-1}a}H_{a_{0}\cdots a_{p-1}}}{p!(s+t+u+1/2)}
\left[k_{2a}(t+1/4)(\alpha'\xi.k_{3})
-
\frac{1}{2}\xi_a(s+1/4)(t+1/4)+(3\leftrightarrow 2)\right]\nonumber\\
&&\times \sum_{n,m=0}^{\infty}d_{n,m}(s+t+1/2)^n((t+1/4)(s+1/4))^m\labell{masslesspole1}\eeqa
where some of the coefficients $d_{n,m}$ are\beqa
d_{0,0}=-\pi^2/3,\,&&d_{1,0}=8\z(3)\nonumber\\
d_{2,0}=-7\pi^4/45,\, d_{0,1}=\pi^4/45,\,&&\,d_{3,0}=32\z(5),\, d_{1,1}=-32\z(5)+8\z(3)\pi^2/3\nonumber\eeqa
The  Mandelstam variables in \reef{masslesspole1} are
\beqa
s=-\frac{\alpha'}{2}(k_1+k_3)^2,&& t=-\frac{\alpha'}{2}(k_1+k_2),\,\,\,\, u=-\frac{\alpha'}{2}(k_2+k_3)^2\eeqa
In above,  $k_1$ is momentum of the  gauge field and $k_2, k_3$ are the tachyon momenta. It has been speculated in \cite{Garousi:2007si} that the above massless poles should be reproduced by the higher derivative theory once one knows the two-tachyon-two-gauge field couplings.

Now using the two-tachyon-two-gauge field couplings \reef{ddbarcoupling}, one can reproduce the string theory massless poles \reef{masslesspole1}. To this end,  consider  the amplitude for decaying   one RR field  to two tachyons and one gauge field in the world-volume theory of brane-anti-brane which has the following  Feynman diagrams:  
\begin{center}
\begin{picture}
(600,100)(0,0)
\Line(25,105)(75,70)\Text(50,105)[]{$T_{1}$}
\Line(25,70)(75,70)\Text(45,80)[]{$T_{1}$}
\Photon(25,35)(75,70){4}{7.5}\Text(50,39)[]{$A^{(1)}$}
\Photon(75,70)(125,70){4}{7.5}\Text(105,88)[]{$A$}
\Gluon(125,70)(175,105){4.20}{5}\Text(145,105)[]{$C_{p-1}$}
\SetColor{Black}
\Vertex(75,70){1.5} \Vertex(125,70){1.5}
\Text(105,20)[]{(a)}
\Line(295,105)(345,70)\Text(320,105)[]{$T_{1}$}
\Line(295,70)(345,70)\Text(320,80)[]{$T_1$}
\Photon(295,35)(345,70){4}{7.5}\Text(320,35)[]{$A^{(1)}$}
\Gluon(345,70)(395,105){4.20}{5}\Text(365,105)[]{$C_{p-1}$}
\Vertex(345,70){1.5}
\Text(345,20)[]{(b)}
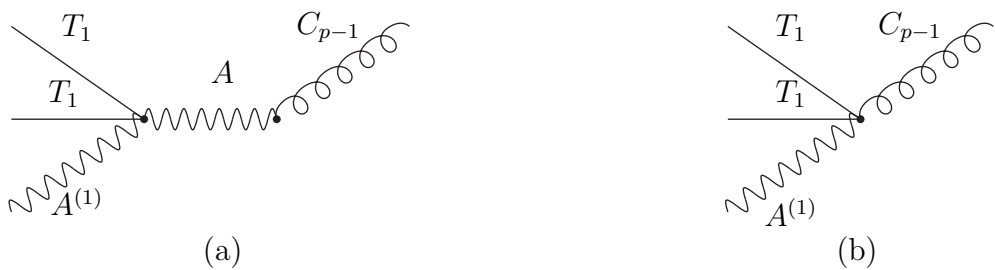
\end{picture}\\ {\sl Figure 3 : The Feynman diagrams corresponding to the amplitude in \reef{amp4}.} 
\end{center}
In field theory, this amplitude is given by
\beqa
{\cal A}&=&V_a(C_{p-1},A)G_{ab}(A)V_b(A,T_1,T_1,A^{(1)})\labell{amp4}\eeqa
where  $A$ should  be $A^{(1)}$ and $A^{(2)}$. In above $T_1$ is the real component of the complex tachyon, \ie $T=(T_1+iT_2)/\sqrt{2}$. It has been  argued in \cite{Garousi:2007si} that the coupling between one RR and one gauge field which is given by the Wess-Zumino terms, and the kinetic term of the gauge field have  no higher derivative correction. Hence, they are given by \cite{Garousi:2007fk} 
\beqa
G_{ab}(A) &=&\frac{i\delta_{ab}}{(2\pi\alpha')^2 T_p
\left(-\alpha' k^2/2\right)}\nonumber\\
V_a(C_{p-1},A^{(1)})&=&i\mu_p(2\pi\alpha')\frac{1}{p!}\epsilon_{a_0\cdots a_{p-1}a}H^{a_0\cdots a_{p-1}}\labell{amp4'}\\
V_a(C_{p-1},A^{(2)})&=&-i\mu_p(2\pi\alpha')\frac{1}{p!}\epsilon_{a_0\cdots a_{p-1}a}H^{a_0\cdots a_{p-1}}\nonumber
\eeqa
The vertexes $ V_b(A,A^{(1)},T_1,T_1)$ which  have higher derivative corrections, can be  derived from  \reef{ddbarcoupling}. One finds that the vertex $V_b(A^{(1)},A^{(1)},T_1,T_1)$ is given by
\beqa
&&-iT_p(\pi\alpha')(\alpha')^2
(-\alpha')^{n+m}a_{n,m}\left(k_b\left[(s+1/4)(2k_2\cdot\xi)+(t+1/4)(2k_3\cdot\xi)\right]\right.\labell{vA1}\\
&&+\left.\left[k_{2b}(t+1/4)(2\xi.k_{3})
+k_{3b}(s+1/4)(2\xi.k_{2})-\xi_b(s+1/4)(t+1/4)\right]\right)\nonumber\\
&&\times\left(\frac{}{}(k_3\inn k)^m(k_3\inn k_1)^n+(k_3\inn k)^n(k_3\inn k_1)^m+(k_2\inn k)^m(k_2\inn k_1)^n+
(k_2\inn k)^n(k_2\inn k_1)^m\right.\nonumber\\
&&\left.+(k_3\inn k)^m(k_2\inn k)^n+(k_3\inn k)^n(k_2\inn k)^m+(k_1\inn k_2)^m(k_1\inn k_3)^n+
(k_1\inn k_2)^n(k_1\inn k_3)^m\frac{}{}\right)\nonumber\eeqa
and the vertex $V_b(A^{(2)},A^{(1)},T_1,T_1)$ by
\beqa
&&-iT_p(\pi\alpha')(\alpha')^2(-\alpha')^{n+m}b_{n,m}\left(k_b\left[(s+1/4)(2k_2\cdot\xi)+(t+1/4)(2k_3\cdot\xi)\right]\right.\labell{vA2}\\
&&+\left.\left[k_{2b}(t+1/4)(2\xi.k_{3})
+k_{3b}(s+1/4)(2\xi.k_{2})-\xi_b(s+1/4)(t+1/4)\right]\right)\nonumber\\
&&\times\left(\frac{}{}(k_3\inn k)^m(k_3\inn k_1)^n+(k_3\inn k)^n(k_3\inn k_1)^m+(k_2\inn k)^m(k_2\inn k_1)^n+
(k_2\inn k)^n(k_2\inn k_1)^m\right.\nonumber\\
&&\left.+(k_3\inn k)^m(k_2\inn k)^n+(k_3\inn k)^n(k_2\inn k)^m+(k_1\inn k_2)^m(k_1\inn k_3)^n+
(k_1\inn k_2)^n(k_1\inn k_3)^m\frac{}{}\right)\nonumber\eeqa
where  $k^a$ is the momentum of the off-shell gauge field. 
Now one can write $k_1\inn k=-(k_1\inn k_2+k_1\inn k_3)$, $k_2\inn k=k_1\inn k_3-k^2$ and $k_3\inn k=k_1\inn k_2-k^2$. The $k^2$ in the above vertex will be canceled with the $k^2$ in the denominator of the gauge field propagator resulting  a bunch of contact terms of one RR, two tachyons and one gauge field, \ie the diagram (b). They should be subtracted from  the contact terms that have been extracted  from the S-matrix element of one RR, two tachyons and one gauge field, \ie the couplings  in \reef{highpn'}. Let us at the moment  ignore the contact terms and consider only the massless poles of the amplitude \reef{amp4}, \ie diagram (a). Replacing \reef{vA2}, \reef{vA1} and \reef{amp4'} in \reef{amp4}, one finds the following massless pole:
\beqa
&&2i\mu_p(\alpha')^2\frac{ \eps^{a_{0}\cdots a_{p-1}a}H_{a_{0}\cdots a_{p-1}}}{p!(s'+t'+u)}
\left[k_{2a}(t')(\alpha'\xi.k_{3})
-\frac{1}{2}\xi_a(s')(t')+(3\leftrightarrow 2)\right]\labell{masslesspole}\\
&&\times \sum_{n,m=0}^{\infty}\left((a_{n,m}-b_{n,m})((t')^m(s')^n+(t')^n(s')^m)\right)\nonumber\eeqa
where $t'=t+1/4=-\alpha' k_1\inn k_2$ and $s'=s+1/4=-\alpha'k_1\inn k_3$. 

The above amplitude should be compared with the  massless poles in \reef{masslesspole1}. 
Let us compare them for some values of $n,m$. For $n=m=0$, the amplitude \reef{masslesspole} has the following factor:
\beqa
4(a_{0,0}-b_{0,0})&=&-\frac{\pi^2}{3}\nonumber\eeqa
which is equal to $d_{0,0}$. At the  order of $\alpha'$, the amplitude \reef{masslesspole} has the following factor:
\beqa
2(a_{1,0}+a_{0,1}-b_{0,1}-b_{0,1})(s'+t')&=&8\z(3)(s+t+1/2)\nonumber\eeqa
which is equal to $d_{1,0}(s+t+1/2)$ in \reef{masslesspole1}. At the  order of $(\alpha')^2$, the amplitude \reef{masslesspole} has the following factor:
\beqa
&&4(a_{1,1}-b_{1,1})(s')(t')+2(a_{0,2}+a_{2,0}-b_{0,2}-b_{2,0})[(s')^2+(t')^2]\nonumber\\
&&=-\frac{7\pi^4}{45}(s'+t')^2+\frac{\pi^4}{45}(s')(t')=d_{2,0}(s+t+1/2)^2+d_{0,1}(s+1/4)(t+1/4)\nonumber\eeqa
At the  order of $(\alpha')^3$, the amplitude \reef{masslesspole} has the following factor:
\beqa
&&2(a_{3,0}+a_{0,3}-b_{0,3}-b_{3,0})[(s')^3+(t')^3]+2(a_{1,2}+a_{2,1}-b_{1,2}-b_{2,1})(s')(t')(s'+t')\nonumber\\
&&=32\z(5)(s'+t')^3+(-32\z(5)+8\pi^2\z(3)/3)(s')(t')(s'+t')\nonumber\eeqa
which is again exactly equal to the corresponding terms in \reef{masslesspole1}. Similar comparison can be done for all order of $\alpha'$. Hence, the field theory amplitude \reef{masslesspole} reproduces  exactly the infinite tower of the massless pole of string theory amplitude \reef{masslesspole1}. In particular, this consistency requires to have couplings between $F^{(1)}$ and $F^{(2)}$ which is in fact  the case, as the coefficients $b_{n,m}$ in \reef{ddbarcoupling} are non-zero.

Finally, let us now return to  the contact terms that the field theory amplitude \reef{amp4} produces. Using the Binomial formula, one can write the contact terms as  the following:
\beqa
&&-i\mu_p(\alpha')^2\frac{ \eps^{a_{0}\cdots a_{p-1}a}H_{a_{0}\cdots a_{p-1}}}{p!}
\left[k_{2a}(t')(\alpha'\xi.k_{3})
-\frac{1}{2}\xi_a(s')(t')+(3\leftrightarrow 2)\right]\sum_{n,m=0}^{\infty}(a_{n,m}-b_{n,m})\nonumber\\
&&\left[\left(2\sum_{\ell=1}^m \pmatrix{m\cr 
\ell}(t'^{m-\ell}s'^n+s'^{m-\ell}t'^n)+2\sum_{\ell=1}^n \pmatrix{n\cr 
\ell}(t'^{n-\ell}s'^m+s'^{n-\ell}t'^m)\right)(\alpha'k^2)^{\ell-1} \right.\nonumber\\
&&\left.+\sum_{\ell=1,j=1}^{n,m} \pmatrix{n\cr 
\ell}\pmatrix{m\cr 
j}(t'^{n-\ell}s'^{m-j}+s'^{n-\ell}t'^{m-j})(\alpha'k^2)^{\ell+j-1}\frac{}{}\right]\labell{masslesspole4}
\eeqa
Note that the above couplings have at least four momenta. They can be rewritten in the following form:
\beqa
&&i(\alpha')^2\mu_p\frac{ \eps^{a_{0}\cdots a_{p-1}a}H_{a_{0}\cdots a_{p-1}}}{p!}
\left[k_{2a}(t+1/4)(\alpha'\xi.k_{3})
-
\frac{1}{2}\xi_a(s+1/4)(t+1/4)+(3\leftrightarrow 2)\right]\nonumber\\
&&\times \sum_{p,n,m=0}^{\infty}e'_{p,n,m}(s+t+u+1/2)^p(s+t+1/2)^n((t+1/4)(s+1/4))^m\eeqa
where $e'_{p,n,m}$ can be written in term of $a_{n,m}$ and $b_{n,m}$. The contact terms of one RR, two tachyons and one gauge field that have been extracted from string theory S-matrix element in \cite{Garousi:2007si}  have the above structure. Hence, the coefficients $e_{p,n,m}$ in the couplings \reef{highpn'}  should be replaced by \beqa
e_{p,n,m}\rightarrow e_{p,n,m}-e'_{p,n,m}\eeqa
This makes the higher derivative  theory  to produce     the string theory  S-matrix element. Since the couplings 
\reef{ddbarcoupling} have on-shell ambiguity, the contact terms in \reef{masslesspole4} have also on-shell ambiguity. The on-shell ambiguity of the couplings in \reef{lagrang} or in  \reef{ddbarcoupling}, however, may be fixed by studying the S-matrix element of three tachyons and two gauge fields because the couplings \reef{lagrang} appear in the tachyon poles and in the contact terms of this S-matrix element.   It would be interesting to perform this calculation.

{\bf Acknowledgement}: H.G.  would like to thank K. Bitaghsir Fadafan   for discussion.

%\newpage

\end{document}